\documentclass[reprint,superscriptaddress,amsmath,amssymb,aps]{revtex4-1}
\usepackage{graphicx} 
\usepackage{float}
\usepackage{color}
\usepackage{textcomp}
\makeatother

\begin{document}

\title{\textcolor{black}{Quantum Simulator} Based on the Paraxial Wave Equation}

\author{Micheline B.~Soley}

\affiliation{Department of Chemistry, University of Wisconsin-Madison, 1101 University
Avenue, Madison, Wisconsin 53706, USA}

\affiliation{Department of Physics, University of Wisconsin-Madison, 1150 University
Avenue, Madison, Wisconsin 53706, USA}

\author{Deniz D.~Yavuz}

\affiliation{Department of Physics, University of Wisconsin-Madison, 1150 University
Avenue, Madison, Wisconsin 53706, USA}

\begin{abstract}

We propose a paraxial quantum simulator that  requires \textcolor{black}{only} widely available optical fibers or  metamaterials. Such a simulator would \textcolor{black}{facilitate} cost-effective quantum simulation without specialized techniques.  We show theoretically that the method accurately simulates quantum dynamics and quantum effects for an example system, 
\textcolor{black}{which invites extension of the method to many-body systems using nonlinear optical elements and} implementation of the 
paraxial quantum simulator to extend access to quantum computation \textcolor{black}{and  prototype quantum parity-time reversal ($\mathcal{PT}$) symmetric technologies}.
    
\end{abstract}

\maketitle

Quantum simulators often require significant investment in specialized hardware, \cite{buluta2009quantum,houck2012chip,bloch2012quantum,schmidt2013circuit,barthelemy2013quantum,georgescu2014quantum,zohar2015quantum,gross2017quantum,schafer2020tools,altman2021quantum,fraxanet2022coming} and even existing optics-based quantum simulators can require precise control \cite{aspuru2012photonic,hartmann2016quantum,angelakis2017quantum}, which can present a barrier to enter the field of quantum information science. Here, we propose an optical system that can serve as a quantum simulator that avoids these specialized techniques by capitalizing on the paraxial approximation to the monochromatic Helmholtz equation.

\begin{figure*}
    \centering
    \includegraphics[width=0.5\textwidth]{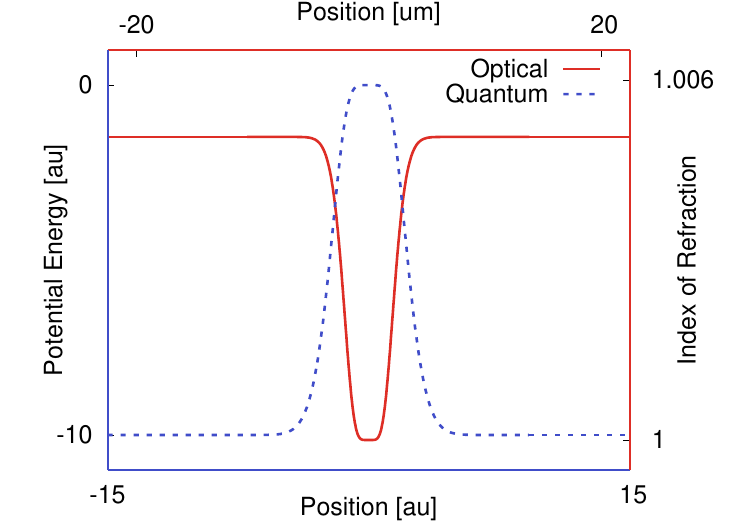} \includegraphics[width=0.45\textwidth]{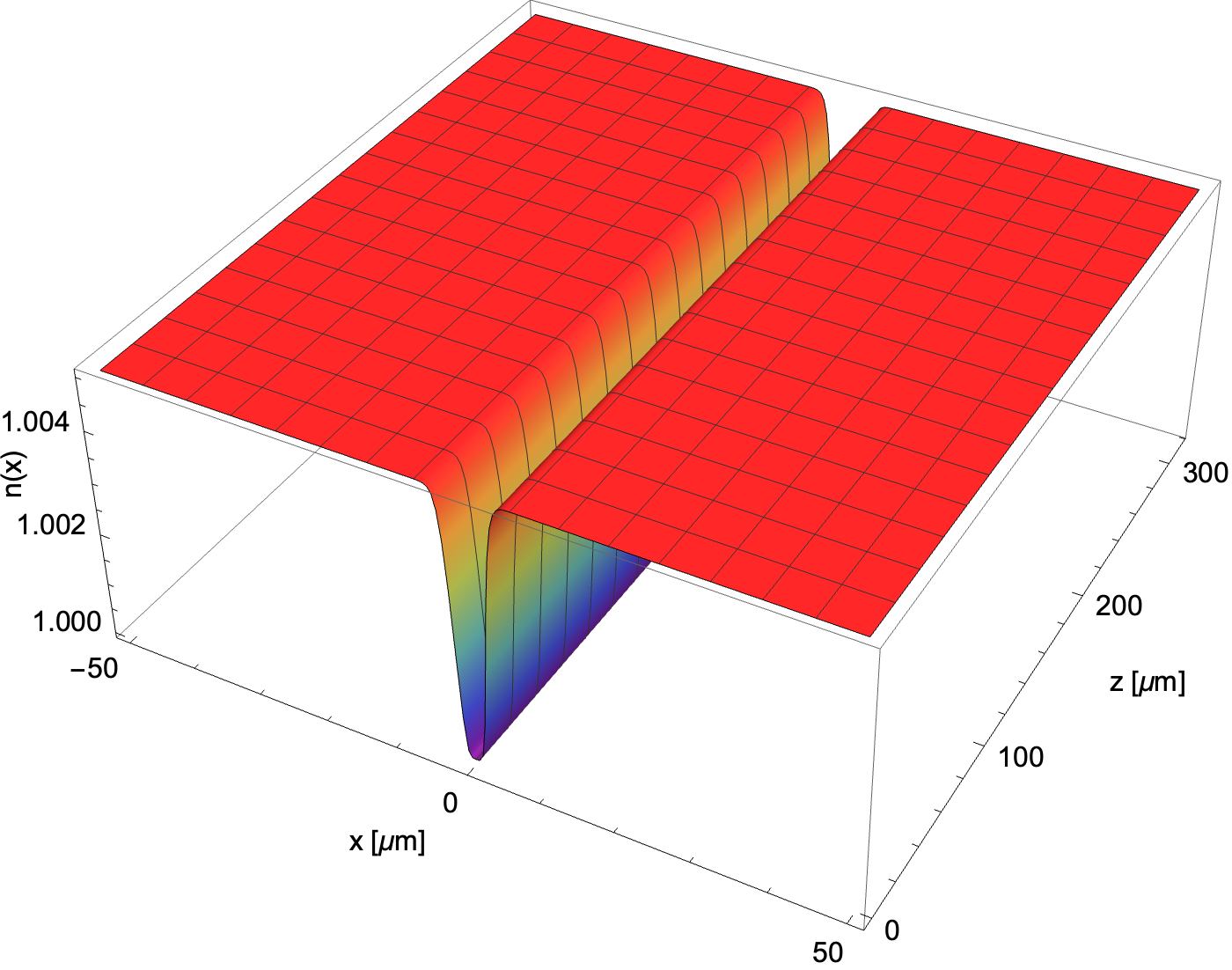}
    \caption{Index of refraction (left) $n(x)$ (solid red line) and (right) $n(x,z)$ (colorful surface) for a paraxial quantum simulator of scattering from the smoothed truncated $V(x)=-x^6$ potential energy surface (dashed blue line). The results correspond to an optical system of position scale $1\text{ \textmu m}$ and wavelength scale \textcolor{black}{$200\text{ nm}$} given a quantum length scale of $1\text{ atomic units}$ and mass of $1\text{ atomic unit}$.}
    \label{fig:potentialindex}
\end{figure*}

\begin{figure*}
    \centering
    \includegraphics[width=\textwidth]{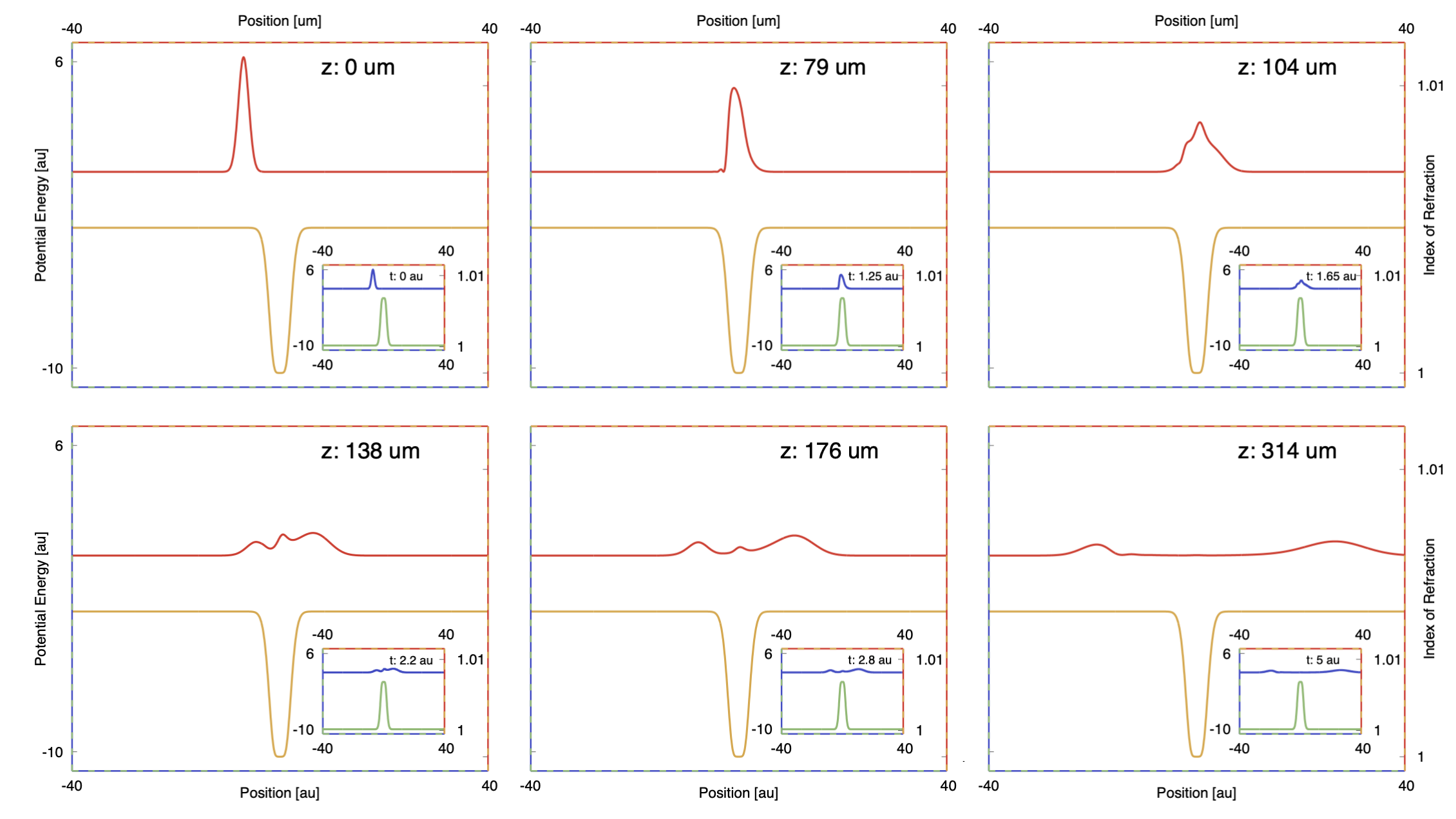}
    \caption{Absolute value of the amplitude squared of propagated light (red line) subject to the index of refraction shown in orange in the paraxial quantum simulator successfully yields the quantum dynamics of the probability density (blue line) subject to the quantum potential shown in green (propagation proceeds from left to right, top to bottom). The absolute value of the amplitude squared and probability density are equivalent except for overall scaling and are vertically scaled and shifted for ease of visualization.}
    \label{fig:dynamics}
\end{figure*}

\textcolor{black}{The paraxial approximation to the wave equation (often referred to as the paraxial wave equation) has been widely used for many decades to study both linear and nonlinear optical phenomenon. The success of the paraxial wave equation is mostly due to the fact that paraxial approximation is well satisfied under most experimental circumstances. This approximation is valid when the divergence half-angles of the wave is smaller than $\sim 0.1$~radians (or equivalently the smallest spatial feature size of the optical system is larger than $\sim 4 \lambda $). }
Although to our knowledge the paraxial wave equation has not been used as a general-purpose ``exact" quantum simulator, it has had immense success as a way to observe optical analogs of a wide array quantum effects \cite{de1989transverse,marte1997paraxial,schwartz2007transport,longhi2009quantum,angelakis2017quantum} including parity-time reversal ($\mathcal{PT}$) symmetry \cite{el2007theory,musslimani2008optical,makris2008beam,longhi2009quantum,ruter2010observation,konotop2016nonlinear,longhi2018parity,ozdemir2019parity}, which suggests that the method could be broadly applicable to a range of quantum phenomena if reformulated for quantum simulation\textcolor{black}{, including the possibility of many-body systems through the introduction of nonlinear optical elements. This would enable this classical, optical method to provide a cost-effective form of analog quantum simulation \cite{cirac2012goals,Hangleiter2022} with intriguing connections to photonic quantum computers \cite{aspuru2012photonic}.}

To provide the basis for such a ``paraxial quantum simulator," we first identify a mapping between quantum and optical variables using the known formal equivalence between the paraxial approximation to the monochromatic Helmholtz equation
\begin{equation}
\text{i}\frac{\partial U(x,z)}{\partial z}=-\frac{1}{2k}\frac{\partial^2 U(x,z)}{\partial x^2}+\textcolor{black}{\frac{1}{2}k}\left[1-n^2(x)\right]U(x,z) \label{eq:TimeDependentParaxialFirst}
\end{equation}
for positions $x$ and $z$, \textcolor{black}{optical} $k$-vector $k$, index of refraction $n(x)$, and amplitude $U(x,z)$ \textcolor{black}{(either the electric field or the magnetic field of the wave\textcolor{black}{, which may be a coherent superposition state})} and the time-dependent Schr{\"o}dinger equation
\begin{equation}
    \text{i}\hbar\frac{\partial \psi(x,t)}{\partial t}=-\textcolor{black}{\frac{\hbar^2}{2m}}\frac{\partial^2 \psi(x,t)}{\partial x^2}+V(x)\psi(x,t)\nonumber
\end{equation}
for position $x$, time $t$, mass $m$, potential $V(x)$, wavefunction $\psi(x)$, and Planck's constant $\hbar$. As previously recognized, this formal equivalence clarifies that the two equations are analogous, provided a mapping between $k$ and $m$, $x$ and $x$, and $t$ and $z$. What has made direct quantum simulation difficult up to now is the missing experimentally realizable optical equivalent to the quantum  potential $V$ and the energy $E$.

We address the difficulty as follows. First, we identify an exact mapping between the optical index of refraction $n(x)$ and the quantum potential $V(x)$. Typically, the index of refraction in optical analog systems is chosen not to yield a specific quantum potential, but rather to demonstrate a specific physical phenomena (for example, to demonstrate the analogue of quantum $\mathcal{PT}$ symmetry in optics, one generates a $\mathcal{PT}$-symmetric index of refraction that is the same under combined complex conjugation and $x\rightarrow -x$ substitution). \textcolor{black}{Noting the term-by-term similarity between the above two equations}, we instead choose the index of refraction to be
\begin{equation}
    n(x)=\sqrt{1-\frac{2}{k}V(x)}\label{eq:indexrefraction}
\end{equation}
such that light subject to $n(x)$ behaves equivalently to a wavepacket subject to $V(x)$, given a suitable mapping between all remaining physical quantities. This index of refraction is realizable with standard optical techniques as long as $n(x)$ remains close to one, which holds as long as $V(x)$ is negative and has a sufficiently small energy bound and both $k$ and $V(x)$ are scaled appropriately \textcolor{black}{ and as long as the rate of change of the potential with position remains within the resolution of the optical fabrication method}. Second, we develop an optical parallel for the quantum energy $E$. It has long been recognized that there is no native, non-derived-quantity optical parallel of the quantum energy in the standard ``time-dependent" form of the paraxial approximation \footnote{Note there have been alternative treatments of the optical axial wave vector as the analog of energy in analyses of $\mathcal{PT}$ symmetry [48, 49].}. We recognize that an exact optical parallel for $E$ appears when one reexpresses the paraxial approximation in ``time-independent" form (see Supplemental Material [SM] Sec.~II for full derivation \cite{SupplementalMaterial}). 
Substitution of the amplitude $U(x,z)=U(x)\exp{(\text{i}\kappa z)}$ into the standard form of the paraxial approximation yields 
\begin{equation}
    -\frac{1}{2k}\frac{\text{d}^2 U(x)}{\text{d}x^2}+V_\text{eff}(x)U(x)=\kappa U(x)\nonumber,
\end{equation}
where $V_\text{eff}(x)=\frac{1}{2}k\left[1-n^2(x)\right]$, which is analogous to the time-independent Schr{\"o}dinger equation
\begin{equation}
    -\frac{\hbar^2}{2m}\frac{\text{d}^2 \psi(x)}{\text{d}x^2}+V(x)\psi(x)=E \psi(x)\nonumber
\end{equation}
given a suitable mapping between $k$ and $m$, $x$ and $x$, and $\kappa$ and $E$. This form emphasizes that, just as the energy $E$ corresponds to a \textcolor{black}{temporal} Fourier component of a propagated wavepacket in quantum scattering calculations, the eigenvalue $\kappa$ corresponds to a measurable \textcolor{black}{spatial} Fourier component of a propagated wavepacket in optical scattering experiments.

We take advantage of this identified equivalency between the paraxial approximation and the Schr{\"o}dinger equation (see further details in SM Sec.~I \cite{SupplementalMaterial}) to simulate one-dimensional quantum scattering as follows.  Initially, we send a monochromatic beam of light in the form of a Gaussian wavepacket to one side of the scatterer near $x_{\min}$ and $z=z_{\min}=0$. \textcolor{black}{This forms the boundary value for the paraxial wave equation of Eq.~(1), i.e., $U(x, z=0)$. We then numerically propagate the boundary value using Eq.~(1) and calculate the amplitude $U(x, z)$ for all $0< z < z_{\max}$, up to a final propagation length $z=z_{\max}$. Because of the above-mentioned equivalence between the two equations, this corresponds to simulating the quantum dynamics of a wave-packet that corresponds to a final time $t_{\max}$ for optical wavenumber scale $w$}. Finally, we compute the reflection coefficient at each energy $E$ through Fourier transformation of the final wavepacket at the energy $E$ and calculation of the ratio between the integral of the probability density in the reflected region $x<0$ relative to the integral of the probability density over all position space $x\in[ x_{\min},x_{\max} ]$, which is possible via measurement of the angle of the scattered light.

To demonstrate the power of the technique, we theoretically model paraxial quantum simulation of scattering in the truncated upside-down $V(x)=-x^6$
 potential \cite{soley2023experimentally}. Scattering from this potential is important in the field of non-Hermitian physics, as reflectionless scattering modes of the truncated potential yield the energies of surprising bound states in the continuum in the infinite-length, unbounded, upside-down $\mathcal{PT}$-symmetric $V(x)=-x^6$ potential system \cite{bender1998real,bender2007making, bender2018scattering,soley2023experimentally}. \textcolor{black}{(See SM Sec.~III for additional results for the truncated upside-down $V(x)=-x^2$ potential \cite{SupplementalMaterial}, for which a paucity of reflectionless scattering modes at low energy agrees with the known absence of bound states in the continuum in the corresponding infinite-length $V(x)=-x^2$ potential \cite{kemble1935contribution,Bender.1998.5243,Bender.2007.947,soley2023experimentally}.)}

We begin by mapping the desired quantum potential $V(x)$ to the optical index of refraction $n(x)$ of the paraxial quantum simulator according to Eq.~\eqref{eq:indexrefraction}. Since the paraxial approximation only holds for sufficiently smooth potentials, we consider a smoothed version of the truncated $V(x)=-x^6$ potential shown in Fig.~\ref{fig:potentialindex}, which has a leading term of the Fourier series at the origin that is sixth-order and an asymptotic behavior that is constant, as desired. The position scale of the scatterer $x$ gives the breadth of the paraxial simulator $x\rightarrow sx$, and the time for the wavepacket to pass through the potential $t$ gives the length of the paraxial simulator $z\rightarrow 2ws^2 t$ for optical wavenumber scale $w$ and wavelength scale $s$. 

To highlight the accessibility of the paraxial quantum simulator, we present theoretical calculations for an optical system with physical dimensions in the micron regime (here one micron) and wavelengths in the \textcolor{black}{ultraviolet regime (here $200$ nm)}, which is readily achievable on a variety of optical platforms. The required index of refraction for the aforementioned parameter choices is shown in Fig.~\ref{fig:potentialindex}. The index of refraction is well within reach of existing experimental methods, as it contains features that are well-approximated by digital lithography techniques with submicrometer accuracy and reaches only a moderate index of refraction of $1.005$ \cite{zheludev2012metamaterials,kadic20193d}. Note that since the paraxial quantum simulator makes no assumptions about the characteristics of the optical system other than that it satisfies the paraxial approximation, an even broader range of physical dimension and wavelength scales are possible, as long as the maximum index of refraction
\begin{equation}
    n_{\max}=\sqrt{1+\frac{10}{s^2w^2}}\nonumber
\end{equation}
remains physically achievable with the material under consideration \textcolor{black}{(see SM Sec.~IV for derivation \cite{SupplementalMaterial})}.

The theoretical calculations indicate the paraxial quantum simulator reproduces quantum dynamics in the potential exactly, \textcolor{black}{where the paraxial approximation holds}, as shown in Fig.~\ref{fig:dynamics}\textcolor{black}{, where calculations for both systems are performed using the standard grid-based split-operator Fourier transform method (which, given the aforementioned equivalence of the Schr{\"o}dinger and paraxial wave equations, differs only by a pair of scaling factors)}.  In both the optical system and the quantum system, a 
\textcolor{black}{Gaussian wavepacket corresponding to a quantum particle of mass $m=0.5\text{ au}$, central position $x_0=-7\text{ au}$, central momentum $p_0=3.5\text{ au}$, and standard deviation $\sigma=\sqrt{2}\text{ au}$} sent towards the scatterer from the left interacts with the scatterer and bifurcates into a reflected wavepacket and a transmitted wavepacket.

\begin{figure}
    \centering
    \includegraphics[width=\columnwidth]{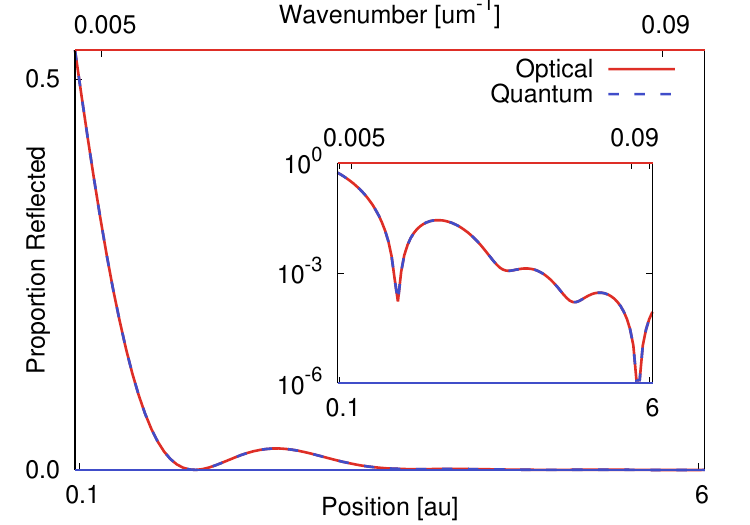}
    \caption{Reflection coefficient computed via the theoretical optical paraxial quantum simulator (solid red line) and split-operator Fourier transform quantum dynamics (dashed blue line) agree exactly over orders of magnitude of the reflection coefficient, modulo a predetermined scaling factor.}
    \label{fig:reflection}
\end{figure}

The paraxial quantum simulator also exactly reproduces the quantum reflection coefficient \textcolor{black}{where the paraxial approximation holds}, as shown in Fig.~\ref{fig:reflection}.
The lowest-energy reflectionless scattering mode is readily visible near $E=1\text{ au}$/$\lambda=0.1\text{  \textmu m}$. And, the optical and quantum results agree over nearly six orders of magnitude in the reflection coefficient. Since the \textcolor{black}{quantum effect }of reflectionless scattering modes correspond to weakly bound states in $\mathcal{PT}$-symmetric systems \cite{ahmed2005reflectionless,soley2023experimentally}, to our knowledge, this constitutes the first direct time-dependent \textcolor{black}{approach to weakly-bound-state (bound-state-in-the-continuum) energy determination} in purely real quantum $\mathcal{PT}$-symmetric systems. \textcolor{black}{We note that it is relatively straightforward to experimentally investigate the theoretical prediction of Fig.~\ref{fig:reflection}. With the given refractive index profile, one can use angle-resolved spectroscopy to measure the reflection and transmission coefficients for incident ``wavepackets," similar to what is depicted in Fig.~2. }

Previously, quantum studies focused on the paraxial approximation as a way to identify optical analogs of quantum effects \cite{de1989transverse,marte1997paraxial,schwartz2007transport,longhi2009quantum,angelakis2017quantum} such as $\mathcal{PT}$-symmetry behavior \cite{el2007theory,musslimani2008optical,makris2008beam,longhi2009quantum,ruter2010observation,konotop2016nonlinear,longhi2018parity,ozdemir2019parity}, but were not able to simulate quantum effects directly. This work recognizes that the technique proposed here, combined with the recent discovery of a class of physically realizable, purely real $\mathcal{PT}$-symmetric systems \cite{soley2023experimentally}, makes exact simulation of quantum effects such as $\mathcal{PT}$-symmetry behaviors possible with the paraxial approximation. This provides both a way to prototype certain quantum $\mathcal{PT}$ technologies and to unique create ``twinned" $\mathcal{PT}$ technologies, such as filters and sensors, based on equivalent same mechanisms on optical and quantum platforms.

Finally, since the paraxial quantum simulator makes no assumptions about the identity of the particles under study, it is broadly applicable to quantum systems. For example, for decades the quantum chaos community sought techniques to use the paraxial approximation to model phenomena such as quantum scars \cite{prange1989experimental,sundaram1995chaos,sundaram1999wave,lemos2012experimental,keski2019quantum}, which would now be possible exactly with the paraxial quantum simulation technique. Likewise, the ultracold \textcolor{black}{atom and molecule} community faces the problem that the cost of exact quantum mechanics simulations on classical computers precludes investigation of all but the smallest systems \cite{balakrishnan2016perspective,croft2017long,bause2023ultracold}. Generalization of the proposed technique \textcolor{black}{to many-body systems through the use of optical nonlinearities} could enable simulation of these low-energy \textcolor{black}{systems} instead on optical platforms already readily accessible to the optics community, which invites new perspectives on quantum simulation applications and the development of new techniques to advance the broader field of quantum computation. \textcolor{white}{\cite{schindler2011experimental,Lin.2011.213901}}

{\em Acknowledgments.} The authors thank  R.~H.~Goldsmith, A.~D.~Stone, \textcolor{black}{ and J. Feinberg} for stimulating discussions. \textcolor{black}{Support for this research was provided by the Office of the Vice Chancellor for Research and Graduate Education at the University of Wisconsin-Madison with funding from the Wisconsin Alumni Research Foundation.}

\bibliography{paraxialbib}

\end{document}


\title{Supplemental Material: \textcolor{black}{Quantum Simulator} Based on the Paraxial Wave Equation}

\author{Micheline B.~Soley}

\affiliation{Department of Chemistry, University of Wisconsin-Madison, 1101 University
Avenue, Madison, Wisconsin 53706, USA}

\affiliation{Department of Physics, University of Wisconsin-Madison, 1150 University
Avenue, Madison, Wisconsin 53706, USA}

\author{Deniz D.~Yavuz}

\affiliation{Department of Physics, University of Wisconsin-Madison, 1150 University
Avenue, Madison, Wisconsin 53706, USA}

\maketitle

\onecolumngrid

\section{Derivation of the Standard Form of the Paraxial Approximation}

To derive the standard form of the paraxial approximation, one begins with the monochromatic Helmholtz equation for light propagating in a medium (for example, a waveguide) with refractive index $n(x)$
\begin{equation}
    0=\frac{\partial^2 E(x,z)}{\partial z^2}+\frac{\partial^2 E(x,z)}{\partial x^2}+k^2 n^2(x) E(x,z)=0. \label{eq:Helmholtz}
\end{equation}
Here, $z$ is the overall propagation direction and for simplicity we only consider one transverse axis $x$. We also assume the refractive index to be only a function of $x$ (namely, there is no variation of the refractive index as a function of $z$). The quantity $k$ is the optical $k$-vector where $k=\omega/c$ for angular frequency $\omega$ and speed of light $c$.

We then focus on waves that predominantly propagate along the $z$ axis and make the paraxial approximation that the light remains near the optical axis throughout. For this purpose, we take
\begin{equation}
E(x,z)=U(x,z)\exp(\text{i}kz), \label{eq:PropagatingLight}
\end{equation}
where the variation of the amplitude $U(x,z)$ along the $z$ axis with reference to the $k$-vector satisfies
\begin{equation}
    \left|\frac{\partial U(x,z)}{\partial z}\right| \ll \left| k\frac{\partial U(x,z)}{\partial z} \right|\nonumber.
\end{equation}
Without any approximation, substitution of Eq.~\eqref{eq:PropagatingLight} into Eq.~\eqref{eq:Helmholtz} gives
\begin{align}
    0&=\frac{\partial^2 U(x,z)}{\partial z^2}+2\text{i}k\frac{\partial U(x,z)}{\partial z}+\frac{\partial^2 U(x,z)}{\partial x^2}+k^2\left[n^2(x)-1\right]U(x,z) \label{eq:PropagatingLightinHelmholtz}.
\end{align}

We now make the paraxial approximation (also frequently referred to as the slowly varying envelope approximation [SVEA], slowly varying asymmetric approximation [SVAA], or narrow-band approximation) by ignoring the first term in Eq.~\eqref{eq:PropagatingLightinHelmholtz} to obtain
\begin{align}
    0&=2\text{i}k\frac{\partial U(x,z)}{\partial z}+\frac{\partial^2 U(x,z)}{\partial x^2}+k^2\left[n^2(x)-1\right]U(x,z)\nonumber
\end{align}
or equivalently
\begin{equation}
0=\text{i}\frac{\partial U(x,z)}{\partial z}+\frac{1}{2k}\frac{\partial^2 U(x,z)}{\partial x^2}-V_\text{eff}(x)U(x,z) \label{eq:TimeDependentParaxial}.
\end{equation}
Here the quantity $V_\text{eff}(x)$ can be thought of as the effective potential and is given by
\begin{equation}
    V_\text{eff}(x)=\frac{1}{2}k\left[1-n^2(x)\right]\nonumber.
\end{equation}

Note Eq.~\eqref{eq:TimeDependentParaxial} has an identical form to the time-dependent Schr{\"o}dinger equation in atomic units with $\left(x,z,k\right)\rightarrow\left(x,t,m,\hbar\right)$ substitution. As a result, one engineers a purely real refractive index shaped in such a way that $V_\text{eff}(x)$ has the appropriate functional dependence on $x$, a wave subject to the paraxial approximation to the monochromatic Helmholtz equation propagates identically to a wavepacket subject to the time-dependent Schr{\"o}dinger equation.

\subsection{Analogy to the time-dependent Schr{\"o}dinger equation}

To highlight the analogy with the time-dependent Schr{\"o}dinger more explicitly, one can perform a change of variables to the standard paraxial approximation expression Eq.~\eqref{eq:TimeDependentParaxial} in which one employs a psuedotime coordinate $\tau=z/c$. Note that here $\tau$ is not meant to be a real time axis, but rather a change of variables such that the expression still describes a monochromatic wave and no variation in time. The resulting expression of the paraxial approximation yields an equation with exactly the same units at the time-dependent Schr{\"o}dinger equation, and therefore establishes an exact analogy between the two expressions. With this change of variables, the standard paraxial approximation expression Eq.~\eqref{eq:TimeDependentParaxial} reads
\begin{equation}
    0=\text{i}\frac{\partial U(x,\tau)}{\partial \tau}-\frac{c}{2k}\frac{\partial^2 U(x,\tau)}{\partial x^2}+cV_\text{eff}(x)U(x,\tau)\nonumber
\end{equation}
or equivalently
\begin{equation}
    0=\text{i}\frac{\partial U(x,\tau)}{\partial \tau}-\frac{c^2}{2\omega}\frac{\partial^2 U(x,\tau)}{\partial x^2}+c V_\text{eff}(x)U(x,\tau)\nonumber.
\end{equation}
We now compare this expression with the time-dependent Schr{\"o}dinger equation
\begin{equation}
    0=\text{i}\hbar\frac{\partial \psi(x,t)}{\partial t}+\frac{\hbar}{2m}\frac{\partial^2 \psi(x,t)}{\partial x^2}-V(x)\psi(x,t)\nonumber.
\end{equation}
The above equations now have exactly the same units and are identical if we choose the frequency of the optical wave to be $\omega=mc^2/\hbar$ and the refractive index profile such that $cV_\text{eff}(x)=V(x)$. The analogy then succeeds experimentally as long as the frequency remains realizable and the index profile is practical using realistic optical parameters.

\section{Derivation of the ``Time-Independent" Paraxial Approximation}

In order to identify the optical parallel of the quantum energy in the standard paraxial approximation expression Eq.~\ref{eq:TimeDependentParaxial}, we seek amplitude solutions of the form $U(x,z)=U(x)\exp(\text{i}\kappa z)$. This solution converts Eq.~\ref{eq:TimeDependentParaxial} to the ``time-independent" equation
\begin{equation}
    -\frac{1}{2k}\frac{\text{d}^2 U(x)}{\text{d}x^2}+V_\text{eff}(x)U(x)=\kappa U(x)\nonumber.
\end{equation}
This is, again, now very similar to the time-independent Schr{\"o}dinger equation, and a full equivalence can be established by multiplying with the speed of light $c$. Note that to consider forward- and backward- propagating waves in quantum scattering equations, we consider solutions of the form $U(x)=\exp(\text{i}k_\text{x}x)$ and $U(x)=\exp(\text{i}k_\text{x}x)$. Without the potential ($V_\text{eff}=0$, namely a ``free particle") these solutions have the dispersion relationship $k_x=\sqrt{2k\kappa}$ in direct analogy to the quantum scattering dispersion relation $k=\sqrt{2mE}$. This formalizes both the direct connection between the quantum energy $E$ and the optical eigenvalue $\kappa$ and the direct connection between the quantum wavenumber $k$ and the optical wavevector $k_x$.

\begin{figure}
    \centering
    \includegraphics[width=0.5\textwidth]{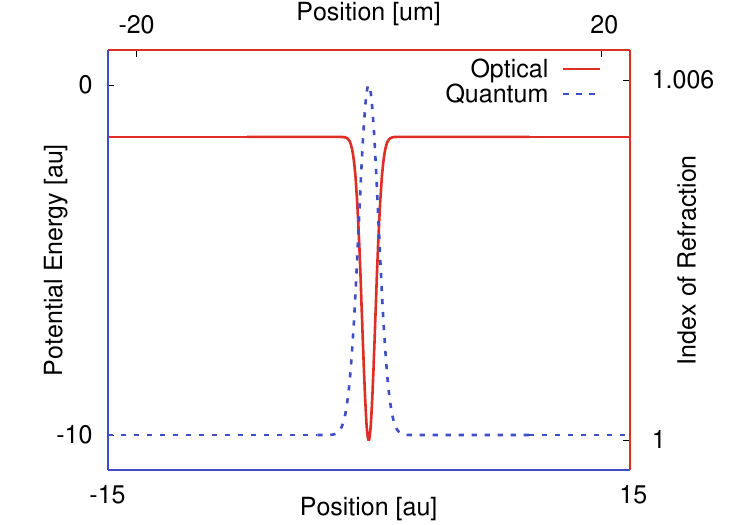} 
    \caption{Index of refraction (left) $n(x)$ (solid red line) required for paraxial quantum simulation of quantum scattering \textcolor{black}{in the smoothed truncated $V(x)=-x^2$ potential} energy surface (dashed blue line) for an optical system of wavelength scale $200\text{ nm}$ and position scale $1\text{ \textmu m}$.}
    \label{fig:200nmpotentialindex}
\end{figure}

\section{Results for Truncated $V(x)=-x^2$ Potential}

\begin{figure*}
    \centering
    \includegraphics[width=1\textwidth]{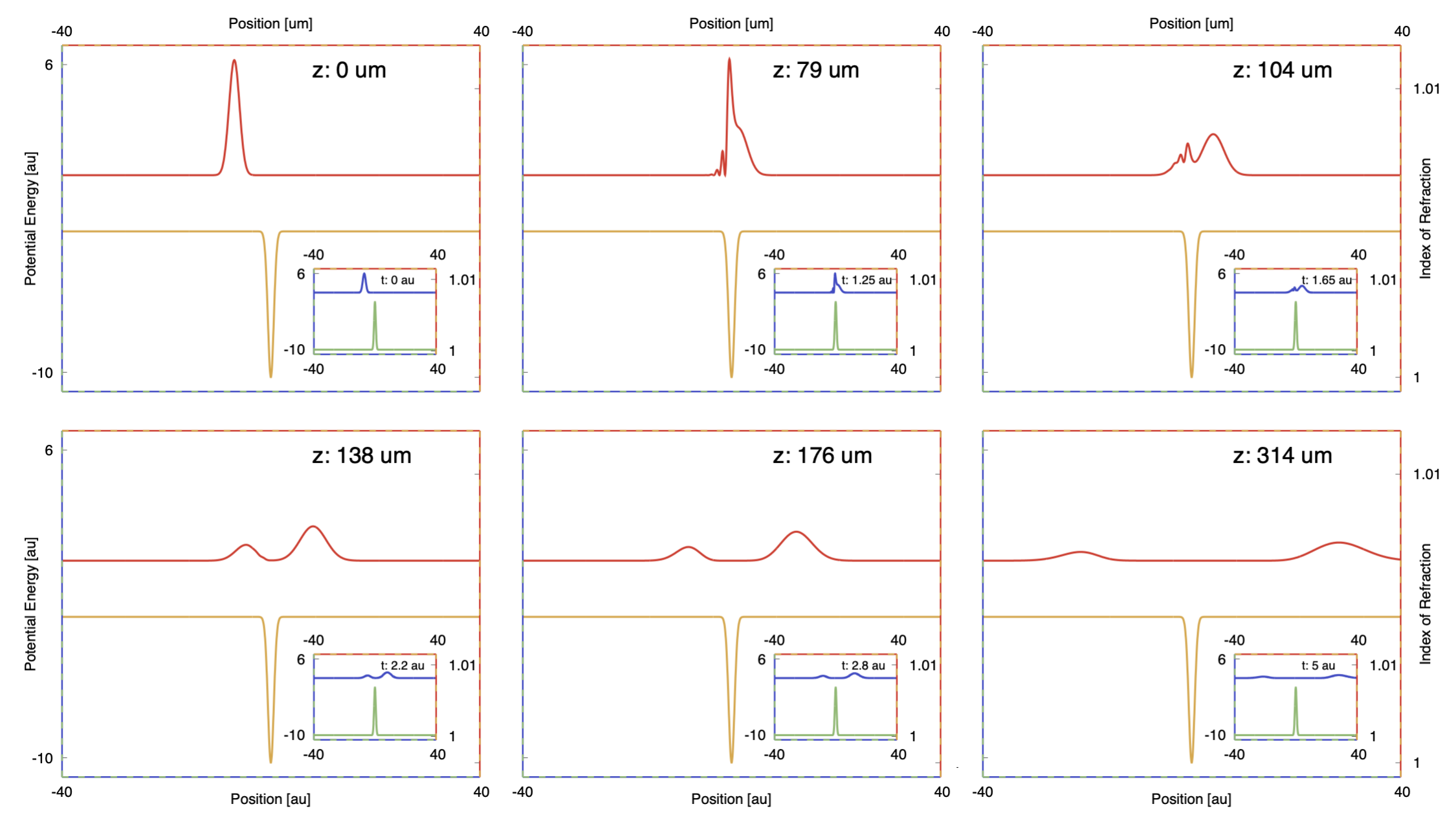}
    \caption{\textcolor{black}{Theoretical results indicate a paraxial quantum simulator successfully reproduces quantum dynamics in the  smoothed truncated $V(x)=-x^2$ potential.}}
    \label{fig:200nmdynamics}
\end{figure*}

\begin{figure}
    \centering\includegraphics[width=0.5\columnwidth]{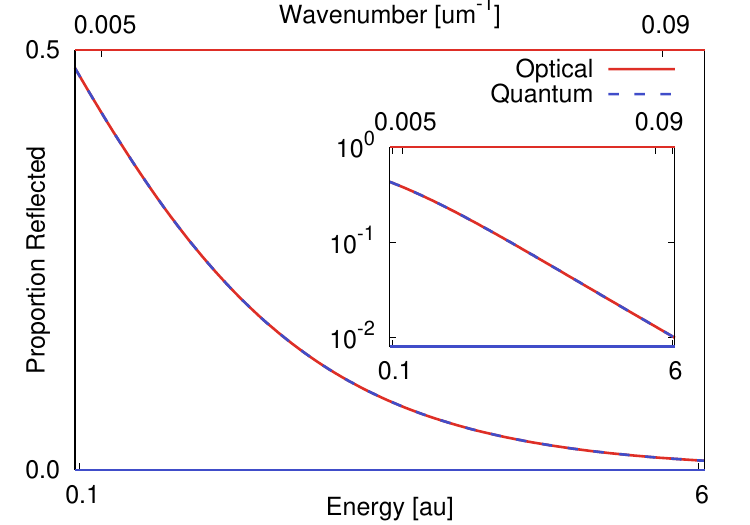}
    \caption{Theoretical reflection coefficient from the paraxial quantum simulator (solid red line) accurately reproduces the quantum result \textcolor{black}{for the smoothed truncated $V(x)=-x^2$ potential} (dashed blue line).}
    \label{fig:200nmreflection}
\end{figure}

We emphasize the \textcolor{black}{broad applicability} the approach by demonstrating theoretically the paraxial optical simulator also successfully simulates quantum scattering in the smoothed truncated \textcolor{black}{$V(x)=-x^2$} potential for a wavelength scale in the ultraviolet range $200\text{ nm}$. As shown in Fig.~\ref{fig:200nmpotentialindex}, the paraxial quantum simulator \textcolor{black}{again} only requires an index of refraction that varies from one by no more than $O(10^{-3})$, which remains well within experimental capabilities. As in the \textcolor{black}{$V(x)=-x^6$} example shown in the main text, the system faithfully reproduces both the dynamics and the reflection coefficient (see Fig.~\ref{fig:200nmdynamics} and Fig.~\ref{fig:200nmreflection}).  \textcolor{black}{As expected, there are no nonmonotonic dips in the reflection coefficient at low energy, in agreement with the fact that the infinite-length $V(x)=-x^2$ supports no bound states in the continuum \cite{kemble1935contribution,Bender.1998.5243,Bender.2007.947,soley2023experimentally}.}

\section{Derivation of Maximum Index of Refraction}

\textcolor{black}{To determine the maximum index of refraction  $n_{\max}$, we first determine the {\em minimum}  potential energy $V_{\min}$. 
Substitution of $V_{\min}$, scaled to chosen length scale $x\rightarrow sx$ and wavelength scale 
$k\rightarrow 2w$, into the formula for the index of refraction then yields
\begin{equation}
    n_{\max}=\sqrt{1-\frac{1}{s^2w^2}V_{\min}}.
\end{equation}
For the particular case of a smoothed truncated $V(x)=-x^6$ potental, which reaches a minimum value of $V_{\min}=-10\text{ au}$ in the asymptotic region, the maximum index of refraction is therefore $n_{\max}=\sqrt{1+10/(s^2w^2)}$ as in main text.}

\bibliography{paraxialbib}